# Maxwell's equations as a special case of deformation of a solid lattice in Euler's coordinates

G. Gremaud
Institute of Condensed Matter Physics
Swiss Federal Institute of Technology of Lausanne
CH-1015 Lausanne, Switzerland

**Abstract**

It is shown that the set of equations known as "Maxwell's equations" perfectly describe two very different systems: *(1)* the usual electromagnetic phenomena in vacuum or in the matter and *(2)* the deformation of isotropic solid lattices, containing topological defects as dislocations and disclinations, in the case of constant and homogenous expansion. The analogy between these two physical systems is complete, as it is not restricted to one of the two Maxwell's equation couples in the vacuum, but generalized to the two equation couples as well as to *the diverse phenomena of dielectric polarization* and *magnetization of matter*, just as to *the electrical charges* and *the electrical currents.* The eulerian approach of the solid lattice developed here includes Maxwell's equations as a special case, since it stems from a tensor theory, which is reduced to a vector one by contraction on the tensor indices. Considering the tensor aspect of the eulerian solid lattice deformation theory, the analogy can be extended to other physical phenomena than electromagnetism, a point which is shortly discussed at the end of the paper.

## 1 - Introduction

To represent the deformation of solids, one generally uses Lagrange's coordinates to describe the evolution of the deformations, and diverse differential geometries to describe the topological defects contained in the solid.
*The use of Lagrange's coordinates* presents a number of inherent difficulties. From the mathematical point of view, the tensors describing the continuous solid deformation are always of order higher than one concerning the spatial derivatives of the displacement field components, which leads to a very complicated mathematical formalism when the solid presents strong distortions (deformations and rotations). To these mathematical difficulties are added physical difficulties when one has to introduce some known properties of solids. Indeed, the Lagrange's coordinates become practically unusable, for example when one has to describe the temporal evolution of the microscopic structure of a solid lattice (phase transitions) and of its structural defects (point defects, dislocations, disclinations, boundaries, etc.), or when it is necessary to introduce some physical properties of the medium (thermal, electrical, magnetic or chemical properties) leading to scalar, vector or tensor fields in the real space.
*The use of differential geometries* in order to introduce topological defects as dislocations in a deformable continuous medium has been initiated by the work of Nye [1] (1953), who showed for the first time the link between the dislocation density tensor and the lattice curvature. On the other hand, Kondo [2] and Bilby [3] (1954) showed independently that the dislocations can be identified as a crystalline version of the Cartan's concept [4] of torsion of a continuum. This approach was generalized in details by Kröner [5] (1960). However, the use of differential geometries in order to describe the deformable media leads very quickly to difficulties similar to those of the Lagrange's coordinates system. A first difficulty arises from the complexity of the mathematical formalism, which is similar to the formalism of general relativity, what makes very difficult to handle and to interpret the obtained general field equations. A second difficulty arises with the differential geometries when one has to introduce topological defects other than dislocations. For example, Kröner [6] (1980) has proposed that the existence of extrinsic point defects could be considered as extra-matter and introduced in the same manner that matter in general relativity under the form of Einstein equations, which would lead to a pure Riemann's's differential geometry in the absence of dislocations. He has also proposed that the intrinsic point defects (vacancies and interstitials) could be approached as a non-metric part of an affine connection. Finally, he has also envisaged introducing other topological defects, as disclinations for example, by using higher order geometries much more complex, as Finsler or Kawaguchi geometries. In fact, the introduction of differential geometries implies generally a heavy mathematical artillery (metric tensor and Christoffel's symbols) in order to describe the spatiotemporal evolution in infinitesimal local frames, as shown for example in the mathematical theory of dislocations of Zorawski [7] (1967).

In view of the complexity of calculations in the case of Lagrange's coordinates as well as in the case of differential geometries, a much simpler approach of deformable solids, but at least equally rigorous, has been developed on the basis of *the Euler's coordinates* [8].

Regarding *the description of defects (topological singularities),* which can appear within a solid, as dislocations and disclinations, it is a domain of physics initiated principally by the idea of macroscopic defects of Volterra [9] (1907). This domain experienced a rapid development during the twentieth century, as well illustrated by Hirth [10] (1985). The lattice dislocation theory started up in 1934, when Orowan [11], Polanyi [12] and Taylor [13] published independently papers describing the edge dislocation. In 1939, Burgers [14] described the screw and mixed dislocations. And finally in 1956, Hirsch, Horne et Whelan [15] and Bollmann [16] observed independently dislocations in metals by using electronic microscopes. Concerning the disclinations, it is in 1904 that Lehmann [17] observed them in molecular crystals, and in 1922 that Friedel [18] gave them a physical explanation. From the second part of the century, the physics of lattice defects has grown considerably.

In the theory of solid lattice deformation in Euler's coordinates [8], the dislocations and the disclinations can be approached by introducing intuitively the concept of dislocation charges, by using the famous Volterra pipes [19] (1907) and an analogy with the electrical charges. The concept of dislocation charge density appears then in an *equation of geometro-compatibility* of the solid, when the concept of flux of charges is introduced in an *equation of geometro-kinetics* of the solid.

The *rigorous formulation of the charge concept* [8] in the solid lattices makes the essential originality of this approach of the topological singularities. The detailed development of this con-cept leads to the appearance of tensor charges of first order, *the dislocation charges,* associated with *the plastic distortions* of the solid (plastic deformations and rotations), and of tensor charges of second order, *the disclination charges*, associated with *the plastic contortions* of the solid (plastic flexions and torsions). It appears that these topological singularities are quantified in a solid lattice and that they have to appear as *strings (thin tubes),* which can be modelized as unidimensional lines of dislocation or disclination, with *linear charges* $\vec{\Lambda}_i, \vec{\Lambda}$ and $\Lambda$. For a given dislocation, the scalar linear charge $\Lambda$ characterizes the screw part of the dislocation and the vector linear charge $\vec{\Lambda}$ characterizes the edge part of the dislocation (figure 1), and both together define completely the nature of the dislocation, without needing a convention at the contrary of the classical definition of a dislocation with its Burger vector. The topological singularities can also appear as *membranes (thin sheets)* in the lattice, which can be modelized as two-dimensional boundaries of flexion, torsion or accommodation, with *surface charges* $\vec{\Pi}_i, \vec{\Pi}$ and $\Pi$.

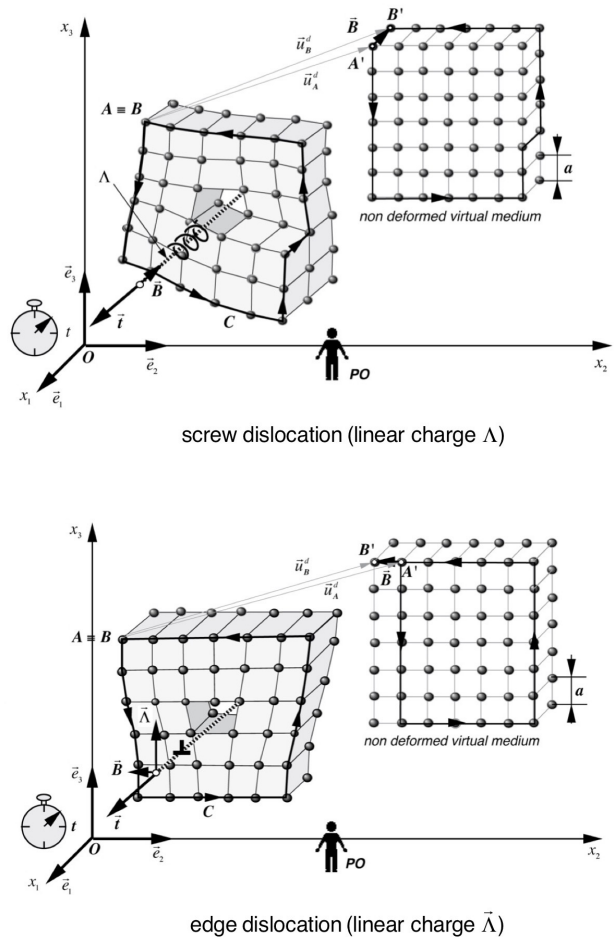

screw dislocation (linear charge $\Lambda$)

edge dislocation (linear charge $\vec{\Lambda}$)

***Figure 1*** *- screw and edge dislocations defined by their respective linear charge* $\Lambda$ *and* $\vec{\Lambda}$

The concept of dislocation and disclination charges allows one to find rigorously the main results obtained by the classical dislocation theory. But it allows above all to define *a tensor density* $\vec{\lambda}_i$ *of dislocation charges*, from which one deduces *a scalar density* $\lambda$ *of rotation charges*, which is associated with the screw part of the dislocations, and *a vector density* $\vec{\lambda}$ *of flexion* *charges*, which is associated with the edge part of the dislocations. On the other hand, the description of the dislocations in the Euler's coordinates by the concept of dislocation charges allows one to treat exactly the evolution of the charges and the deformations of the solid lattice during any kind of volume contractions or expansions, independently if it is small or strong.

As will be shown, such an eulerian approach of the solid lattices allows one to find a perfect and complete analogy between the non-divergent deformation of an isotropic solid lattice and the Maxwell's equations of electromagnetism.

## 2 - Description of the solid lattice deformations in Euler's coordinates

In Euler's coordinates, the solid distortions (local deformations and rotations) can be completely described by *a second order tensor of distortion* $\beta_{ij}$ [8], which can be represented for convenience as a *field of 3 vectors* $\vec{\beta}_i$, by writing

$$\begin{pmatrix} \vec{\beta}_1 \\ \vec{\beta}_2 \\ \vec{\beta}_3 \end{pmatrix} = \begin{pmatrix} \beta_{11} & \beta_{12} & \beta_{13} \\ \beta_{21} & \beta_{22} & \beta_{23} \\ \beta_{31} & \beta_{32} & \beta_{33} \end{pmatrix} \begin{pmatrix} \vec{e}_1 \\ \vec{e}_2 \\ \vec{e}_3 \end{pmatrix} \qquad (1)$$

This tensor is defined as *the gradient tensor of the velocity field* $\vec{\phi}(\vec{r},t)$ of the solid lattice in the Euler's coordinates (that is, the velocity of the lattice, seen as a continuum, at a given point and given time), using the following *geometro-kinetic equation*

$$\frac{d\vec{\beta}_i}{dt} = -\vec{J}_i + \overrightarrow{\text{grad}}\,\phi_i \qquad (2)$$

in which $\vec{J}_i$ represents *the tensor flux of dislocation charges* in the solid and $d/dt = \partial/\partial t + (\vec{\phi}\vec{\nabla})$ represents *the material derivative,* meaning the temporal derivative of a quantity observed along the trajectory of the sites of the lattice.

The trace of the geometro-kinetic equation for $\vec{\beta}_i$ is a geometro-kinetic equation for a scalar $\tau$, which describes *the volume expansion* $\tau$ of the solid lattice

$$\frac{d\tau}{dt} = -\sum_k \vec{e}_k \vec{J}_k + \text{div}\,\vec{\phi} = -\frac{S_n}{n} + \text{div}\,\vec{\phi} \qquad (3)$$

and which is related to *the local source* $S_n$ *of lattice sites*, and to *the divergence of the velocity field* $\vec{\phi}(\vec{r},t)$. The expansion $\tau$ depends on *the local density* $n = n(\vec{r},t)$ *of lattice sites* and on *the local volume of one lattice site* $v = v(\vec{r},t) = 1/n(\vec{r},t)$, by relations

$$\tau = -\ln(n/n_0) = \ln(v/v_0) \qquad (4)$$

In this way, the volume expansion $\tau$ is null when there is neither expansion nor contraction of the lattice ($n = n_0, v = v_0$), and becomes positive or negative if there is respectively *an expansion* ($n < n_0, v > v_0$) or *a contraction* ($n > n_0, v < v_0$) of the lattice.

The anti-symmetric part of the geometro-kinetic equation for $\vec{\beta}_i$ represents a geometro-kinetic equation for *the vector $\vec{\omega}$ of local rotation* of the solid lattice

$$\frac{d\vec{\omega}}{dt} = \frac{1}{2}\sum_k \vec{e}_k \wedge \vec{J}_k + \frac{1}{2}\overrightarrow{\text{rot}}\,\vec{\phi} = -\vec{J} + \frac{1}{2}\overrightarrow{\text{rot}}\,\vec{\phi} \qquad (5)$$

which depends on *the curl of the velocity field* $\vec{\phi}(\vec{r},t)$ and on *the vector flux* $\vec{J}$ *of screw dislocation charges* in the solid.

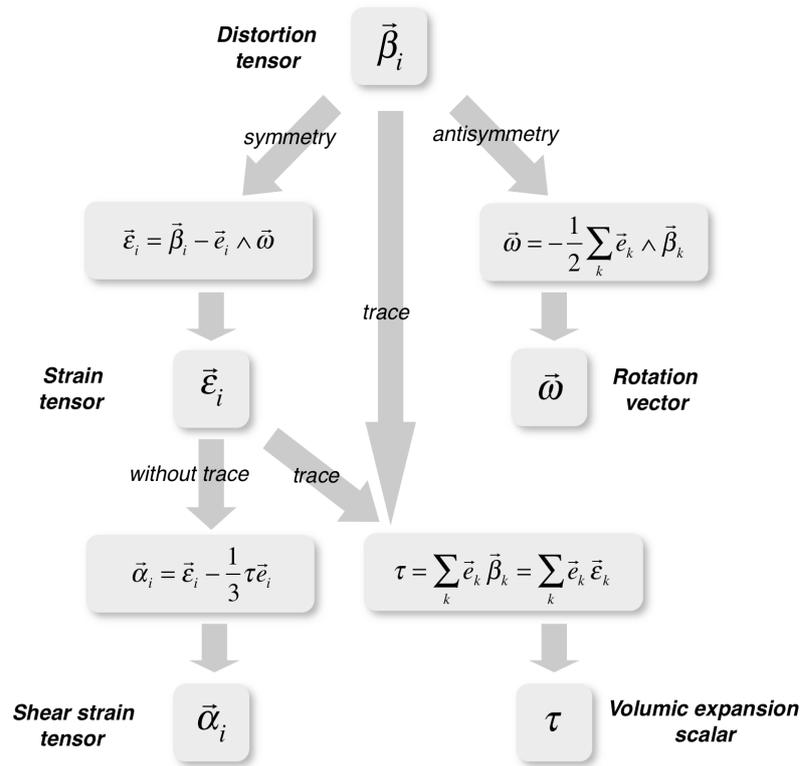

**Figure 2** - Decomposition of tensor of distortion $\vec{\beta}_i$ in Euler's coordinates

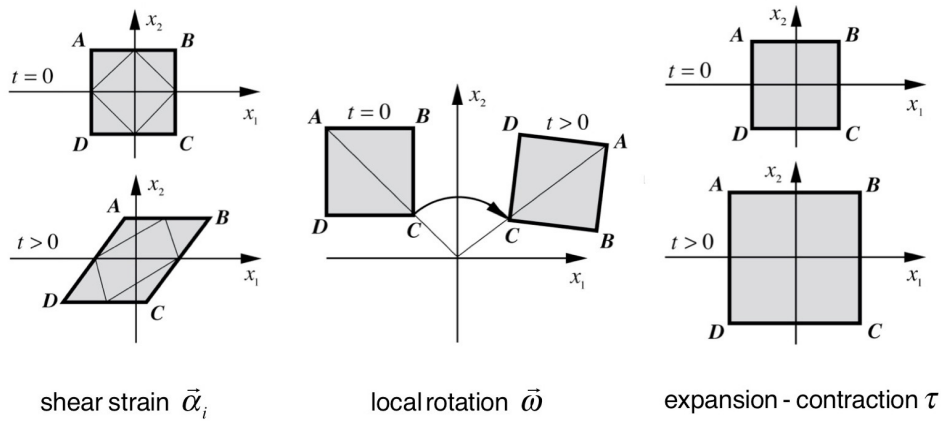

shear strain $\vec{\alpha}_i$     local rotation $\vec{\omega}$     expansion - contraction $\tau$

**Figure 3** - decomposition of distortions as shear strain $\vec{\alpha}_i$, local rotation $\vec{\omega}$ and expansion-contraction $\tau$

The complete decomposition of the tensor of distortion $\vec{\beta}_i$ follows the diagram presented in figure 2. It shows the symmetric and anti-symmetric parts, the trace and the transverse symmetric part of the tensor, which represent respectively *the tensor of deformation* $\vec{\varepsilon}_i$, *the vector of rotation* $\vec{\omega}$, *the scalar of volume expansion* $\tau$ and *the tensor of shear strain* $\vec{\alpha}_i$ of the solid. From this diagram, one deduces that the distortions (deformations and local rotations) of a solid lattice is completely described by giving either $\{\vec{\beta}_i\}$, or $\{\vec{\varepsilon}_i,\vec{\omega}\}$, or $\{\vec{\alpha}_i,\vec{\omega},\tau\}$. This last description can be illustrated as in figure 3.

If the topological fields of distortion $\vec{\beta}_i, \vec{\varepsilon}_i, \vec{\alpha}_i, \vec{\omega}, \tau$ represent the distortions, deformations, shears, rotations and volume expansions that occur at every point and throughout the cells of the lattice, the first and second spatial derivatives of these various topological tensors represent the "*curvatures*" of the solid, as the macroscopic flexions and torsions, and are called the *contortions of the lattice* (figure 4).

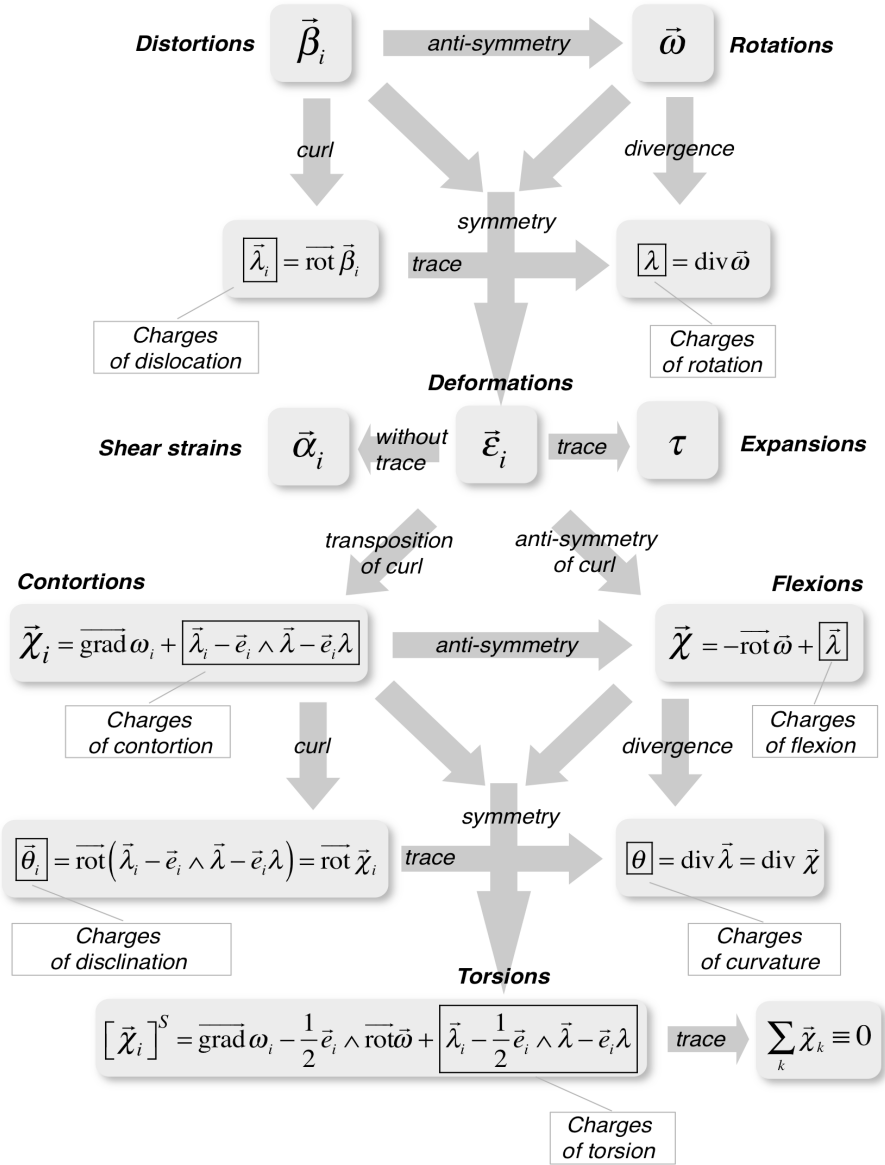

**Figure 4** - contortions and geometro-compatibility equations of a lattice in Euler's coordinates

These contortions are related to *the geometro-compatibility* of the solid, and allows one to introduce the important concepts of *density $\vec{\lambda}_i$ of dislocation charges*, *density $\lambda$ of rotation charges*, *density $\vec{\lambda}$ of flexion charges* and *density $\theta$ of curvature charges*, as illustrated by the diagram of the *geometro-compatibility equations* in figure 4, and to define *the following basic geometro-compatibility equations*

$$\begin{cases} \vec{\lambda}_i = \overrightarrow{\mathrm{rot}}\,\vec{\beta}_i \\ \lambda = \sum \vec{e}_i \vec{\lambda}_i = \mathrm{div}\,\vec{\omega} \\ \vec{\lambda} = \sum \vec{\lambda}_i \wedge \vec{e}_i \quad \Rightarrow \quad \theta = \mathrm{div}\,\vec{\chi} = \mathrm{div}\,\vec{\lambda} \end{cases} \qquad (6)$$

The relations between the flows and densities of charges inside the lattice can be written

$$\begin{cases} \vec{J}_i = \vec{\lambda}_i \wedge \vec{v} \\ \vec{J} = \vec{J}^\lambda + \vec{J}^{\vec{\lambda}} = \lambda \vec{v} + \left(\vec{\lambda} \wedge \vec{v}\right)/2 \\ S_n/n = -\vec{\lambda}\vec{v} \end{cases} \qquad (7)$$

in which $\vec{v}$ represents the *velocity of the charges* with regard to the lattice.
*If there is no creation or annihilation of charges inside the lattice*, the conservation equations for the charges can be written

$$\begin{cases} \dfrac{d\vec{\lambda}_i}{dt} = -\overrightarrow{\mathrm{rot}}\,\vec{J}_i \\ \dfrac{d\vec{\lambda}}{dt} = -2\,\overrightarrow{\mathrm{rot}}\,\vec{J}^{(\vec{\lambda})} - \vec{v}\,\mathrm{div}\,\vec{\lambda} \\ \dfrac{d\lambda}{dt} = -\mathrm{div}\,\vec{J}^{(\lambda)} \end{cases} \qquad (8)$$

### 3 - Eulerian continuity equation for the energy of a solid lattice

The total energy $E$ contained in a volume $V_m$ of solid is equal to the integral on this volume of the sum of *the kinetic energy density* $T_{cin}$ and *the internal energy density* $U$, which is to say

$$E = T + U = \iiint_{V_m} (T_{cin} + U)\,dV \qquad (9)$$

According to the first principle of thermodynamics, this energy is a conserved quantity inside an isolated volume $V_m$. So any variation in said energy can only be due to external influences coming from outside the volume $V_m$, translating the fact that this volume is in fact *non-isolated*. These variations are due either to work exchange or heat exchange. By definition those exchanges either result from a pass through the boundary surface $S_m$ around the volume $V_m$, or result from work due to an external work performed on the internal medium by an outside field (such as gravity, electro-magnetism, etc). The exchange of work and heat across the surface $S_m$ can be represented as *a surface flux of work* $\vec{J}_w$ and *a surface flux of heat* $\vec{J}_q$. Concerning the volume work due to outside forces, it is introduced in the form of *a volume source of work of external forces* $S_w^{ext}$.

The principle of conservation of energy in a given volume $V_m$ can be written as the equality of the temporal derivation of total energy contained in said volume $V_m$ and the different contribu-tions to the volume, which comprises the work volume sources we just discussed and the pass-through contributions, which necessarily cross the boundary surface $S_m$

$$\frac{d}{dt}\iiint_{V_m}(T_{cin}+U)dV = \iiint_{V_m} S_w^{ext}\, dV - \oiint_{S_m}(\vec{J}_w+\vec{J}_q)d\vec{S} \qquad (10)$$

Using the theorem of divergence and the derivation of the integral on the volume $V_m$ mobile with the lattice, this equation can be transformed into the following one

$$\iiint_{V_m}\left[\frac{\partial T_{cin}}{\partial t}+\frac{\partial U}{\partial t}+\text{div}(T_{cin}\vec{\phi}+U\vec{\phi})\right]dV = \iiint_{V_m}\left[S_w^{ext}-\text{div}(\vec{J}_w+\vec{J}_q)\right]dV \qquad (11)$$

which, under local form, gives

$$\frac{\partial T_{cin}}{\partial t}+\frac{\partial U}{\partial t}+\text{div}(T_{cin}\vec{\phi})+\text{div}(U\vec{\phi}) = S_w^{ext}-\text{div}\vec{J}_w-\text{div}\vec{J}_q \qquad (12)$$

It is judicious to transform this equation, which contains volume densities $T_{cin}$ and $U$, by introducing the measures $e_{cin}$ and $u$, defined respectively as the *average kinetic energy* and the *average internal energy,* defined per elementary site of the lattice

$$e_{cin} = T_{cin}/n \quad \text{and} \quad u = U/n \qquad (13)$$

Thanks to these definitions, one obtains the final formulation of the *first principle*, or *principle of continuity of energy* in Euler's coordinates, in its local form

$$n\frac{du}{dt}+n\frac{de_{cin}}{dt} = S_w^{ext}-\text{div}\vec{J}_w-\text{div}\vec{J}_q-uS_n-e_{cin}S_n \qquad (14)$$

## 4 - Newton's equation of a perfect solid lattice

For the simplest solid lattice, which will be called *the perfect solid*, with *a newtonian inertial mass $m$ per site*, the kinetic energy per site is given by

$$e_{cin} = \frac{1}{2}m\vec{\phi}^2 \qquad (15)$$

And for a perfect *isotropic* solid lattice, the elastic deformation part $U^{def}$ of the internal energy per volume unit of the lattice could depend linearly on the volume expansion field $\tau$, quadratically on the volume expansion field $\tau$, on the rotation field $\vec{\omega}$ and on the shear elastic strain field $\vec{\alpha}_i$, via four elastic modulus $K_0, K_1, K_2$ and $K_3$

$$U^{def}(\tau,\vec{\alpha}_i,\vec{\omega}) = -K_0\tau+K_1\tau^2+K_2\sum_i(\vec{\alpha}_i)^2+2K_3(\vec{\omega})^2 \qquad (16)$$

so that the internal energy per site $u$ can be written, introducing the thermal part $u^{th}$ which is a function of *the entropy $s$ per site*

$$u(s,\tau,\vec{\omega},\vec{\alpha}_i) = u^{th}(s)+\frac{1}{n}\left(-K_0\tau+K_1\tau^2+K_2\sum_i(\vec{\alpha}_i)^2+2K_3(\vec{\omega})^2\right) \qquad (17)$$

One can then express the derivative term of this energy

$$n\frac{du}{dt} = n\frac{\partial u}{\partial s}\frac{ds}{dt}+n\frac{\partial u}{\partial \tau}\frac{d\tau}{dt}+n\frac{\partial u}{\partial \vec{\omega}}\frac{d\vec{\omega}}{dt}+n\sum_i\frac{\partial u}{\partial \vec{\alpha}_i}\frac{d\vec{\alpha}_i}{dt} = nT\frac{ds}{dt}-p\frac{d\tau}{dt}+\vec{m}\,\frac{d\vec{\omega}}{dt}+\sum_i\vec{s}_i\frac{d\vec{\alpha}_i}{dt} \qquad (18)$$

which allows one to define *the intensive quantities of temperature $T$, pressure $p$, torque mo-mentum $\vec{m}$ and *shear stress* $\vec{s}_i$ inside the solid lattice, and *the state equations* for these quantities

$$T = \frac{\partial u}{\partial s} \ , \ p = -n\frac{\partial u}{\partial \tau} = K_0 - 2K_1\tau - U^{el} \ , \ \vec{m} = \frac{\partial u}{\partial \vec{\omega}} = 4K_3\vec{\omega} \ , \ \vec{s}_i = \frac{\partial u}{\partial \vec{\alpha}_i} = 2K_2\vec{\alpha}_i \quad (19)$$

With these state equations, relation *(14)* can be rewritten as

$$n\frac{du}{dt} + n\frac{de_{cin}}{dt} = nT\frac{ds}{dt} - p\frac{d\tau}{dt} + \vec{m}\frac{d\vec{\omega}}{dt} + \sum_i \vec{s}_i \frac{d\vec{\alpha}_i}{dt} + nm\vec{\phi}\frac{d\vec{\phi}}{dt}$$
$$= S_w^{ext} - \mathrm{div}\,\vec{J}_w - \mathrm{div}\,\vec{J}_q - uS_n - e_{cin}S_n \quad (20)$$

With respect to the source of work $S_w^{ext}$ due to a field of external forces, it can be calculated if we know the nature of the field force. For example, if we suppose that the solid lattice is in a *constant gravity field* $\vec{g}$, we will have

$$S_w^{ext} = \rho\vec{\phi}\vec{g} = mn\vec{\phi}\vec{g} \quad (21)$$

One can also introduce in *(20)* the *average linear momentum* $\vec{p}$ per site of the solid lattice as

$$\vec{p} = m\vec{\phi} \quad (22)$$

And using the fact that the shear strain $\vec{s}_i$ is *a transverse symmetrical tensor* (transverse means with a null trace), one has

$$\sum_i \vec{s}_i \frac{d\vec{\alpha}_i}{dt} = \sum_i \vec{s}_i \frac{d\vec{\beta}_i}{dt} \quad (23)$$

so that *(20)* can be written

$$nT\frac{ds}{dt} - p\frac{d\tau}{dt} + \vec{m}\frac{d\vec{\omega}}{dt} + \sum_i \vec{s}_i \frac{d\vec{\beta}_i}{dt} + n\vec{\phi}\frac{d\vec{p}}{dt} = \rho\vec{\phi}\vec{g} - \mathrm{div}\,\vec{J}_w - \mathrm{div}\,\vec{J}_q - uS_n - e_{cin}S_n \quad (24)$$

Introducing now the geometro-kinetic equations *(2), (3)* and *(5)* in this relation, one obtains finally, after some transformations

$$\vec{\phi}\left(n\frac{d\vec{p}}{dt} + \overrightarrow{\mathrm{grad}}\,p + \frac{1}{2}\overrightarrow{\mathrm{rot}}\,\vec{m} - \sum_i \vec{e}_i\,\mathrm{div}\,\vec{s}_i\right) - \mathrm{div}\left(-\sum_i \phi_i\vec{s}_i - \frac{1}{2}(\vec{\phi}\wedge\vec{m}) + p\vec{\phi}\right)$$
$$+ nT\frac{ds}{dt} + \left(p\frac{S_n}{n} - \vec{m}\vec{J} - \sum_i \vec{s}_i\,\vec{J}_i\right) = \vec{\phi}(\rho\vec{g}) - \mathrm{div}(\vec{J}_w) - \mathrm{div}\,\vec{J}_q - uS_n - e_{cin}S_n \quad (25)$$

The only way to always satisfy this equation is that the three following equations are still individually satisfied

$$n\frac{d\vec{p}}{dt} = \rho\vec{g} + \sum_k \vec{e}_k\,\mathrm{div}\,\vec{s}_k - \frac{1}{2}\overrightarrow{\mathrm{rot}}\,\vec{m} - \overrightarrow{\mathrm{grad}}\,p \quad (26a)$$

$$\vec{J}_w = -\sum_i \phi_i\vec{s}_i - \frac{1}{2}(\vec{\phi}\wedge\vec{m}) + p\vec{\phi} \quad (26b)$$

$$nT\frac{ds}{dt} = -(u + e_{cin})S_n - \left(p\frac{S_n}{n} - \vec{m}\vec{J} - \sum_i \vec{s}_i\,\vec{J}_i\right) - \mathrm{div}\,\vec{J}_q \quad (26c)$$

The first equation *(26a)* is in fact *the Newton's equation of the solid lattice*. The second one *(26b)* corresponds to the expression of *the surface flux of work* $\vec{J}_w$, which has the same meaning than *the Poynting's vector* in electromagnetism. The third one *(26c)* is *the heat equation of the solid lattice*, which shows that *heat sources and sinks* are activated in the presence of *crea-* tion-annihilation $S_n$ of lattice sites, of flows $\vec{J}$ and $\vec{J}_i$ of charges and of heat flows $\vec{J}_q$ inside the lattice.

In the heat equation *(26c)*, the term $\vec{m}\vec{J} + \sum \vec{s}_i\,\vec{J}_i - pS_n/n$ is in fact *the power $P_{ch}$ supplied to the dislocation charges by the stress fields*. In this power term, it is possible to replace the flows $\vec{J}_i$ and $\vec{J}$ and the source $S_n/n$ by their expressions as a function of the velocity $\vec{v}$ of the charges using relations *(7)*. One obtains after some transformations an expression of $P_{ch}$ containing the charge densities $\vec{\lambda}_k$, $\vec{\lambda}$ and $\lambda$

$$P_{ch} = \left[\sum_k \left(\vec{s}_k \wedge \vec{\lambda}_k\right) + \lambda \vec{m} + \frac{1}{2}\left(\vec{m} \wedge \vec{\lambda}\right) + \vec{\lambda}p\right]\vec{v} \tag{27}$$

The power $P_{ch}$ given to the charges is thus the product of a velocity by a term that is *the force $\vec{f}_{PK}$ acting by volume unit* on the charge densities $\vec{\lambda}_k$, $\vec{\lambda}$, $\lambda$

$$P_{ch} = \vec{f}_{PK}\vec{v} \quad\Rightarrow\quad \vec{f}_{PK} = \sum_k \left(\vec{s}_k \wedge \vec{\lambda}_k\right) + \lambda \vec{m} + \frac{1}{2}\left(\vec{m} \wedge \vec{\lambda}\right) + \vec{\lambda}p \tag{28}$$

This force depends on the stress tensors $\vec{s}_k$, $\vec{m}$ and/or $p$, and is usually called *the force of Peach and Koehler*. The use of the stress tensors $\vec{s}_k$, $\vec{m}$ and/or $p$ depends of course on the choice of how we write the internal energy state function *(16)* of the considered solid lattice.

The equation of Newton *(26a)* can be written using the state functions *(19)*, and also the facts that $\overrightarrow{\mathrm{grad}}\,n = -n\,\overrightarrow{\mathrm{grad}}\,\tau$, $\rho = nm$ and $\sum \vec{e}_k \vec{\alpha}_k \equiv 0$. One obtains then the following version of the Newton's equation

$$n\frac{d\vec{p}}{dt} = nm\vec{g} + 2K_2 \sum_i \vec{e}_i \,\mathrm{div}\,\vec{\alpha}_i - 2K_3 \overrightarrow{\mathrm{rot}}\,\vec{\omega} + 2K_1 \overrightarrow{\mathrm{grad}}\,\tau + \overrightarrow{\mathrm{grad}}\,U^{def}(\tau,\vec{\alpha}_i,\vec{\omega}) \tag{29}$$

But it is still possible to find another formulation of the Newton's equation by using the equations giving *the vector of flexion $\vec{\chi}$* of the solid lattice

$$\vec{\chi} = -\sum_k \vec{e}_k \wedge \overrightarrow{\mathrm{rot}}\,\vec{\alpha}_k - \frac{2}{3}\overrightarrow{\mathrm{grad}}\,\tau = \sum_k \vec{e}_k \,\mathrm{div}\,\vec{\alpha}_k - \frac{2}{3}\overrightarrow{\mathrm{grad}}\,\tau = -\overrightarrow{\mathrm{rot}}\,\vec{\omega} + \vec{\lambda} \tag{30}$$

which allows one to connect the space derivatives of the distortion tensors

$$\sum_k \vec{e}_k \,\mathrm{div}\,\vec{\alpha}_k = \frac{2}{3}\overrightarrow{\mathrm{grad}}\,\tau - \overrightarrow{\mathrm{rot}}\,\vec{\omega} + \vec{\lambda} \tag{31}$$

so that the Newton's equation of the solid lattice becomes, if we neglect now the gravitation acceleration ($\vec{g} = 0$)

$$n\frac{d\vec{p}}{dt} = -2(K_2 + K_3)\overrightarrow{\mathrm{rot}}\,\vec{\omega} + \left(\frac{4}{3}K_2 + 2K_1\right)\overrightarrow{\mathrm{grad}}\,\tau + \overrightarrow{\mathrm{grad}}\,U^{def}(\tau,\vec{\alpha}_i,\vec{\omega}) + 2K_2\vec{\lambda} \tag{32}$$

It is remarkable here that the average linear momentum $\vec{p}$ of each site contains a rotational part $\vec{p}^{\,rot}$ due to $\overrightarrow{\mathrm{rot}}\,\vec{\omega}$ and a divergent part $\vec{p}^{\,div}$ due to $\overrightarrow{\mathrm{grad}}\,\tau$ and $\overrightarrow{\mathrm{grad}}\,U^{def}$. If one supposes that the *density $\vec{\lambda}$ of charges of flexion* contains also *a rotational part $\vec{\lambda}^{rot}$* ($\mathrm{div}\,\vec{\lambda}^{rot} = 0$) and *a divergent part $\vec{\lambda}^{div}$* ($\overrightarrow{\mathrm{rot}}\,\vec{\lambda}^{div} = 0$), one can separate the Newton's equation in two partial Newton's equations by writing

$$n\frac{d\vec{p}^{\,rot}}{dt} = -2(K_2 + K_3)\overrightarrow{\mathrm{rot}}\,\vec{\omega} + 2K_2\vec{\lambda}^{rot} \tag{33a}$$

$$n\frac{d\vec{p}^{\,div}}{dt} = \left(\frac{4}{3}K_2 + 2K_1\right)\overrightarrow{\text{grad}}\,\tau + \overrightarrow{\text{grad}}\,U^{def}(\tau,\vec{\alpha}_i,\vec{\omega}) + 2K_2\vec{\lambda}^{div} \qquad (33b)$$

## 5 - Maxwell's equations at constant and homogeneous volume expansion

Suppose now that the deformation of the perfect solid lattice occurs at constant and homogeneous volume expansion, so that

**Hypothesis 1:** $\quad \tau = cste \neq \tau(\vec{r},t) \quad \Rightarrow \quad n = cste \neq n(\vec{r},t)$ \hfill (34)

With this hypothesis, only the rotational part *(33a)* of the Newton's equation of the lattice remains interesting, and one observes that all the deformations, the rotations as well as the shear strains, are *completely described* by the rotation field $\vec{\omega}$. One can introduce here *a new angular momentum* $\vec{m}'$, which is related at the same time to *the elastic rotation* and to *the elastic shear strain* of the lattice, by writing in the following way *a new elastic state equation*

$$\vec{\omega} = \frac{\vec{m}'}{4(K_2 + K_3)} \qquad (35)$$

which allows one to write the Newton's equation, as $n = cste$, in the form

$$\frac{d(n\vec{p}^{\,rot})}{dt} = -\frac{1}{2}\overrightarrow{\text{rot}}\,\vec{m}' + 2K_2\vec{\lambda}^{rot} \qquad (36)$$

One supposes also that only charges with symmetrical tensor of densities $\vec{\lambda}_i$ exist, which means that all the dislocations are screw dislocations, so that the scalar densities $\lambda$ exist in this lattice, but not vector densities $\vec{\lambda}$

**Hypothesis 2:** $\quad \exists\,\lambda = \lambda(\vec{r},t) \quad$ but $\quad \vec{\lambda}^{rot} = 0$ \hfill (37)

In this case, the relations *(7)* and *(8)* for the charges inside the lattice can be summarized by

$$\vec{J} = \vec{J}^{\lambda} = \lambda \vec{v} \quad,\quad S_n = 0 \quad,\quad \frac{d\lambda}{dt} = -\text{div}\,\vec{J} \qquad (38)$$

It is clear, from equations *(3), (34)* and *(39),* that the constancy of the expansion leads to

$$\text{div}(n\vec{p}^{\,rot}) = n\,\text{div}\,\vec{p}^{\,rot} = nm\,\text{div}\,\vec{\phi}^{\,rot} = 0 \qquad (39)$$

All the equations necessary to describe the deformation of the lattice in this case are

- the geometro-kinetic equation and the compatibility equation of the rotations

$$\begin{cases} -\dfrac{d\vec{\omega}}{dt} + \dfrac{1}{2}\overrightarrow{\text{rot}}\,\vec{\phi}^{\,rot} = \vec{J} \\ \text{div}\,\vec{\omega} = \lambda \end{cases} \qquad (40)$$

- the Newton's equation and the non-divergence of the linear momentum

$$\begin{cases} \dfrac{d(n\vec{p}^{\,rot})}{dt} = -\dfrac{1}{2}\overrightarrow{\text{rot}}\,\vec{m}' \\ \text{div}(n\vec{p}^{\,rot}) = 0 \end{cases} \qquad (41)$$

- the generalized elastic state equation and the relation giving the linear momentum

$$\begin{cases} \vec{\omega} = \dfrac{\vec{m}'}{4(K_2 + K_3)} \\ \vec{p}^{rot} = m\vec{\phi}^{rot} \end{cases} \qquad (42)$$

- the continuity equation for the scalar density of rotation charges

$$\begin{cases} \dfrac{d\lambda}{dt} = -\operatorname{div}\vec{J} \end{cases} \qquad (43)$$

One can also suppose that the global rotation vector $\vec{\omega}$ of the lattice can contain *an elastic part* $\vec{\omega}^{el}$ and *an anelastic part* $\vec{\omega}^{an}$, meaning a part which is instantaneous and a part which does not react instantaneously to a change of *the shear stress* $\vec{s}_i$ or of *the momentum* $\vec{m}$. This case is simply taken into account by writing the following relationship

**Hypothesis 3:** $\quad \vec{\omega} = \vec{\omega}^{el} + \vec{\omega}^{an} = \dfrac{\vec{m}'}{4(K_2 + K_3)} + \vec{\omega}^{an}$ \hfill (44)

One can also suppose that the lattice contains some *point defects*, *vacancies and interstitials*, with constant and homogeneous concentrations

**Hypothesis 4:** $\quad C_L = n_L/n = cste \qquad \text{and} \qquad C_I = n_I/n = cste$ \hfill (45)

In this case, one can define an *average effective mass* $m'$ of the lattice sites

$$m' = m(1 + C_I - C_L) = cste \qquad (46)$$

and one can also introduce a *mass transportation by diffusion* in the lattice, by introducing a *non-divergent surface flow* $\vec{J}_L^{rot}$ of vacancies and a *non-divergent surface flow* $\vec{J}_I^{rot}$ of interstitials

**Hypothesis 5:** $\quad \exists \quad \vec{J}_L^{rot} \neq 0 \quad \text{and} \quad \vec{J}_I^{rot} \neq 0 \quad \text{with} \quad \operatorname{div}\vec{J}_L^{rot} = \operatorname{div}\vec{J}_I^{rot} = 0$ \hfill (47)

The *average linear momentum* $\vec{p}$ *per site* of the lattice becomes in this case

$$\vec{p}^{rot} = m'\vec{\phi}^{rot} + m\left(\dfrac{\vec{J}_I^{rot}}{n} - \dfrac{\vec{J}_L^{rot}}{n}\right) = m\left[\vec{\phi}^{rot} + (C_I - C_L)\vec{\phi}^{rot} + \dfrac{1}{n}\left(\vec{J}_I^{rot} - \vec{J}_L^{rot}\right)\right] \qquad (48)$$

It is easy to prove that this momentum is non-divergent, because the expansion of the lattice is constant and homogeneous, such as $\operatorname{div}\vec{\phi}^{rot} = 0$, and the mass transportation by diffusion is related to non-divergent flows by hypothesis 5

$$\operatorname{div}\vec{p}^{rot} = m'\operatorname{div}\vec{\phi}^{rot} + \dfrac{m}{n}\left(\operatorname{div}\vec{J}_I^{rot} - \operatorname{div}\vec{J}_L^{rot}\right) = 0 \quad \Rightarrow \quad \operatorname{div}n\vec{p}^{rot} = 0 \qquad (49)$$

These relations allow one to change the couple of the elastic state equation and of the expression of the linear momentum, without modifying the other equations of the solid lattice

$$\begin{cases} \vec{\omega} = \vec{\omega}^{el} + \vec{\omega}^{an} = \dfrac{\vec{m}'}{4(K_2 + K_3)} + \vec{\omega}^{an} \\ \vec{p}^{rot} = m\left[\vec{\phi}^{rot} + (C_I - C_L)\vec{\phi}^{rot} + \dfrac{1}{n}\left(\vec{J}_I^{rot} - \vec{J}_L^{rot}\right)\right] \end{cases} \qquad (50)$$

We can still establish a *continuity equation for the energy* from equations *(40)* and *(41)*

$$-\vec{m}'\vec{J} = \vec{m}'\frac{d\vec{\omega}}{dt} + \vec{\phi}^{rot}\frac{d(n\vec{p}^{rot})}{dt} - \text{div}\left(\frac{1}{2}\vec{\phi}^{rot} \wedge \vec{m}'\right) \quad (51)$$

It is also easy to show that the transversal waves of rotation and shear strain presents a propagation celerity given by

$$c_t = \sqrt{\frac{K_2 + K_3}{nm}} \quad (52)$$

One can then write all the equations allowing to describe the deformation of this perfect elastic, anelastic and self-diffusive solid lattice at constant and homogeneous volume expansion, as shown in table 1. If the equations are *locally described in a mobile reference frame* $O'x'y'z'$ *in translation with the lattice at velocity* $\vec{\phi}$, the material derivative $d/dt$ representing the temporal variation of a quantity observed along the trajectory of the sites of the lattice can be replaced by the partial derivative $\partial/\partial t$ of this quantity. All the necessary equations are written in table 1, in such a way that they can be compared with the Maxwell's equations of electromagnetism.

**Table 1 - Maxwell's formulation of the equations of a perfect solid lattice presenting homogeneous expansion $\tau$ in the mobile frame $O'x'y'z'$**

$$\begin{cases} -\dfrac{\partial \vec{\omega}}{\partial t} + \overrightarrow{\text{rot}}\,\dfrac{\vec{\phi}^{rot}}{2} = \vec{J} \\ \text{div}\,\vec{\omega} = \lambda \end{cases} \quad\Leftrightarrow\quad \begin{cases} -\dfrac{\partial \vec{D}}{\partial t} + \overrightarrow{\text{rot}}\,\vec{H} = \vec{j} \\ \text{div}\,\vec{D} = \rho \end{cases}$$

$$\begin{cases} \dfrac{\partial n\vec{p}^{rot}}{\partial t} = -\overrightarrow{\text{rot}}\,\dfrac{\vec{m}'}{2} \\ \text{div}\,n\vec{p}^{rot} = 0 \end{cases} \quad\Leftrightarrow\quad \begin{cases} \dfrac{\partial \vec{B}}{\partial t} = -\overrightarrow{\text{rot}}\,\vec{E} \\ \text{div}\,\vec{B} = 0 \end{cases}$$

$$\begin{cases} \vec{\omega} = \left(\dfrac{1}{2(K_2+K_3)}\right)\dfrac{\vec{m}'}{2} + \vec{\omega}^{an} \\ n\vec{p}^{rot} = 2nm\left[\dfrac{\vec{\phi}^{rot}}{2} + (C_I - C_L)\dfrac{\vec{\phi}^{rot}}{2} + \left(\dfrac{1}{2n}(\vec{J}_I^{rot} - \vec{J}_L^{rot})\right)\right] \end{cases} \Leftrightarrow \begin{cases} \vec{D} = \varepsilon_0 \vec{E} + \vec{P} \\ \vec{B} = \mu_0\left[\vec{H} + (\chi^{para} + \chi^{dia})\vec{H} + \vec{M}\right] \end{cases}$$

$$\begin{cases} \dfrac{\partial \lambda}{\partial t} = -\text{div}\,\vec{J} \end{cases} \quad\Leftrightarrow\quad \begin{cases} \dfrac{\partial \rho}{\partial t} = -\text{div}\,\vec{j} \end{cases}$$

$$\begin{cases} -\dfrac{\vec{m}'}{2}\vec{J} = \\ \dfrac{\vec{\phi}^{rot}}{2}\dfrac{\partial n\vec{p}^{rot}}{\partial t} + \dfrac{\vec{m}'}{2}\dfrac{\partial \vec{\omega}}{\partial t} - \text{div}\left(\dfrac{\vec{\phi}^{rot}}{2} \wedge \dfrac{\vec{m}'}{2}\right) \end{cases} \Leftrightarrow \begin{cases} -\vec{E}\vec{j} = \\ \vec{H}\dfrac{\partial \vec{B}}{\partial t} + \vec{E}\dfrac{\partial \vec{D}}{\partial t} - \text{div}(\vec{H} \wedge \vec{E}) \end{cases}$$

$$\begin{cases} c_t = \sqrt{\dfrac{K_2+K_3}{nm}} \end{cases} \quad\Leftrightarrow\quad \begin{cases} c = \sqrt{\dfrac{1}{\varepsilon_0 \mu_0}} \end{cases}$$

## 6 - The analogy between the solid lattice non-divergent deformations and the Maxwell's equations of electromagnetism

The analogy between the equations of the isotropic solid lattice deformations taken at constant and homogeneous volume expansion and the Maxwell's equations of electromagnetism is remarkable, because it is absolutely complete, as clearly shown by the equations given in table 1:

- ***analogy between density of rotation charges and density of electrical charges:*** the equations of table 1 show a complete analogy between the scalar density $\lambda$ of rotation charges and the density $\rho$ of electric charges, as well as between the vector flow $\vec{J}$ of rotation charges and the density of electric current $\vec{j}$.

- ***analogy between anelasticity of the lattice and dielectric properties of matter:*** the phenomenon of anelasticity introduced here by the term $\vec{\omega}^{an}$ becomes, in comparison with Maxwell's equations of electromagnetism, analogous to the dielectric polarization in the relationship $\vec{D} = \varepsilon_0 \vec{E} + \vec{P}$, giving the electric displacement $\vec{D}$ versus the electric field $\vec{E}$ and the polarization of matter $\vec{P}$.

This analogy between fields $\vec{\omega}^{an}$ and $\vec{P}$ is very strong since the possible phenomenological behaviors of these two quantities are entirely similar, as relaxational, resonant or hysteretic behaviors. For example, in the case of a pure relaxation, it is possible to connect $\vec{\omega}$ and $\vec{m}'$ by means of a complex modulus, as it is possible to connect $\vec{D}$ and $\vec{E}$ via a similar complex dielectric coefficient in electromagnetism (in fact, a deeper comparison would show that the behaviors associated with thermal activation in the two cases also present analogies).

- ***analogy between mass transportation in the lattice and magnetism of matter:*** as $n\vec{p}^{rot}$ represents both the average linear momentum per unit volume of the solid and the average mass flow of the solid per unit volume, we deduce that the mass flow within the solid is due at the same time to a mass transport $nm\vec{\phi}^{rot}$ with velocity $\vec{\phi}^{rot}$ corresponding to the movement of the lattice, to a second mass transportation $nm(C_I - C_L)\vec{\phi}^{rot}$ at velocity $\vec{\phi}^{rot}$ by the driving motion of the point defects by the lattice and finally to a mass transportation $m(\vec{J}_I^{rot} - \vec{J}_L^{rot})$ due to the phenomenon of self-diffusion of vacancies and interstitials.

Each of these mass transports has an analog in Maxwell's equations of electromagnetism. The mass transport $nm\vec{\phi}^{rot}$ by the lattice is analogous to the term $\mu_0 \vec{H}$ of the *magnetic induction* in vacuum. The mass transport $nm(C_I - C_L)\vec{\phi}^{rot}$ by dragging along the point defects by the lattice perfectly corresponds to the term $\mu_0(\chi^{para} + \chi^{dia})\vec{H}$ of magnetism, wherein the *magnetic susceptibility* is composed of two parts: the *positive paramagnetic susceptibility* $\chi^{para}$, which becomes the analog of the concentration $C_I$ of interstitials, and the *negative diamagnetic susceptibility* $\chi^{dia}$, which is therefore analogous to the concentration of vacancies $C_L$.

Concerning the phenomena of non-divergent self-diffusion by the vacancies and interstitials, we have in these equations the term $m(\vec{J}_I^{rot} - \vec{J}_L^{rot})$, which links the last part of $n\vec{p}^{rot}$ to the rotational velocity fields $\Delta\vec{\varphi}_L^{rot}$ and $\Delta\vec{\varphi}_I^{rot}$ of vacancies and interstitials with regard to the lattice

$$n\vec{p}_{self-diffusion}^{rot} = m(\vec{J}_I^{rot} - \vec{J}_L^{rot}) = mn(C_I \Delta\vec{\varphi}_I^{rot} - C_L \Delta\vec{\varphi}_L^{rot}) \qquad (53)$$

As an example we can imagine a hypothetical lattice in which the vacancies are tightly anchored to the lattice (friction coefficient $B_L \to \infty$), while the interstitials are free to move (friction coefficient $B_I = 0$). In this case, one can simply write

$$\Delta\vec{\varphi}_L^{rot} \to 0 \quad \text{and} \quad \Delta\vec{\varphi}_I^{rot} \neq 0 \tag{54}$$

As a consequence, the quantity of movement $n\vec{p}^{rot}$ within the lattice can be written

$$n\vec{p}^{rot} = nm\left[\vec{\phi}^{rot} + (C_I - C_L)\vec{\phi}^{rot} + C_I \Delta\vec{\varphi}_I^{rot}\right] \tag{55}$$

Mass transport $n\vec{p}^{rot}$ now has a term $(C_I - C_L)\vec{\phi}^{rot}$ associated with both vacancies and interstitials, whose coefficient $(C_I - C_L)$ is analogous to the magnetic susceptibility $\chi$ in electromagnetism, and that can take a positive or negative value depending on concentrations $C_I$ and $C_L$ of point defects. It further contains the term $nmC_I\Delta\vec{\varphi}_I^{rot}$ associated with the mass transport by inertial conservative interstitial movement, which is perfectly analogous to the permanent magnetization $\vec{M}$ of the ferromagnetic and antiferromagnetic materials in electromagnetism.

**Table 2 - The complete analogy with the Maxwell's theory of electromagnetism**

| | |
|---|---|
| $\vec{\omega}^{él}$ = vector of elastic shear and local rotation<br>$n\vec{p}^{rot}$ = volume linear momentum of lattice<br>     = mass flow of lattice<br>$\vec{m}'/2$ = generalized torque momentum / 2<br>$\vec{\phi}^{rot}/2$ = local velocity field of the lattice / 2 | $\vec{D}$ = electric field of displacement<br>$\vec{B}$ = magnetic induction field<br><br>$\vec{E}$ = electric field<br>$\vec{H}$ = magnetic field |
| $\vec{J}$ = surface flow of screw dislocations<br>$\lambda$ = density of rotation charges<br>   = density of screw dislocations | $\vec{j}$ = electric current<br>$\rho$ = density of electric charges |
| $1/2(K_2 + K_3) = \dfrac{1}{2(\text{shear modulus} + \text{rotation modulus})}$<br>$2nm = 2 \times$ mass density of the lattice | $\varepsilon_0$ = dielectric permittivity of vacuum<br>$\mu_0$ = magnetic permeability of vacuum |
| $\vec{\omega}^{an}$ = vector of anelastic shear and local rotation<br>$(C_I - C_L)$ = atomic concentrations of interstitials and vacancies<br>           of interstitials and vacancies<br>$(\vec{J}_I^{rot} - \vec{J}_L^{rot})/2n = \dfrac{\text{surface flux of interstitials and vacancies}}{2x \text{ density of lattice sites}}$ | $\vec{P}$ = dielectric polarization of matter<br>$(\chi^{para} + \chi^{dia})$ = paramagnetic and diamagnetic<br>                        susceptibility of matter<br>$\vec{M}$ = magnetization of matter |
| $\vec{\phi}^{rot}/2 \wedge \vec{m}'/2$ = vector of Poynting<br>$c_t = \sqrt{(K_2 + K_3)/nm}$ = speed of shear and rotation<br>                  transvesal waves | $\vec{H} \wedge \vec{E}$ = vector of Poynting<br>$c = \sqrt{1/(\varepsilon_0 \mu_0)}$ = speed of light<br>                in vacuum |

The presence of the non-divergent term $nmC_I\Delta\vec{\varphi}_I^{rot}$ in $n\vec{p}^{rot}$ clearly corresponds to a non-Markovian type of process, since its value must depend on the history of this hypothetical solid lattice. One could imagine for instance that the movement of interstitials is controlled by *a dry type of friction* with the lattice, in which case there would be *a critical force of depinning* for interstitials, which would lead to the emergence of *cycles of hysteresis* of $\Delta\vec{\varphi}_I^{rot}(t)$ as a function of $\vec{\phi}^{rot}(t)$. This would be absolutely similar to the cycles of hysteresis of magnetization $\vec{M}(t)$ as a function of the magnetic field $\vec{H}(t)$ observed in ferromagnetic or antiferromagnetic mate-rials. The complete analogy between the parameters of the non-divergent deformations of a solid lattice and the Maxwell's theory of electromagnetism is reported in table 2.

## 7 - Conceivable generalization of the analogy between the eulerian theory of lattice deformation and other modern physics theories

The existence of a similarity or an analogy between two theories is always a very fruitful and successful thing in physics by the reciprocal contribution of one theory to the other.

It is clear that the analogy with the electromagnetic field theory will enable the use of the theoretical tools developed in field theory, such as the Lorentz transformation or the delayed potential theory for example. In the other direction, the theory of non-divergent deformation developed here presents some aspects, which can be discussed in the frame of the analogy with electromagnetism:

*- on the non-existence of an analogy with magnetic monopoles:* as $n\vec{p}^{rot} = mn\vec{\phi}^{rot}$ represents the mass transport in the lattice, the equation $\text{div}(n\vec{p}^{rot}) = 0$ reflects the fact that the mass is a quantity which is conserved, and that there cannot exist a creation or an annihilation of mass inside the perfect lattice. As part of the analogy with electromagnetism, a relationship $\text{div}(n\vec{p}^{rot}) = cste \neq 0$ would be like a $\text{div}\vec{B} = cste \neq 0$ relationship in electromagnetism. Now this last relationship shows the well-known concept of *magnetic monopoles, particles of unipolar magnetic charges*, suggested by some theories of electromagnetism, but never observed experimentally, and which would therefore be in the analogy with the solid lattice deformation *a localized and continuous source of lattice sites or of point defects* in the lattice corresponding to local and constant mass creation or mass annihilation. As a consequence, an analogy with the magnetic monopoles does not exist in the solid lattice deformation theory.

*- on the possible existence of "vector electric charges" in this analogy :* one can legitimately ask what could be the analogy of the density $\vec{\lambda}^{rot}$ of rotational flexion charges appearing in *(33a)* in the case of the Maxwell's equations. If there were a quantity $\vec{\lambda}^{rot}$ similar in the Maxwell equations, one could hypothetically call it a *density $\vec{\rho}$ of "vector electric charges"*, by postulating the following analogy

$$\vec{\lambda}^{rot} \Leftrightarrow \vec{\rho} \tag{56}$$

The equations of Maxwell would then be written a little differently from the known equations, with an extra term of charge, but not in the equation $\text{div}\vec{B} = 0$ as suggested in the theories of magnetic monopoles, but in the equation $\partial \vec{B}/\partial t = -\overrightarrow{\text{rot}}\vec{E}$, in the following way

$$\begin{cases} \dfrac{\partial n\vec{p}^{rot}}{\partial t} = -\overrightarrow{\text{rot}}\dfrac{\vec{m}'}{2} + 2K_2\vec{\lambda}^{rot} \\ \text{div}\, n\vec{p}^{rot} = 0 \end{cases} \Leftrightarrow \begin{cases} \dfrac{\partial \vec{B}}{\partial t} = -\overrightarrow{\text{rot}}\vec{E} + \kappa\vec{\rho} \\ \text{div}\,\vec{B} = 0 \end{cases} \tag{57}$$

in which $\kappa$ would be a *new electric coefficient*, analogous to the modulus $2K_2$

$$2K_2 \Leftrightarrow \kappa \tag{58}$$

In the static case, if such a vector charge did in fact exist, the equation containing it would be written as

$$\frac{\partial \vec{B}}{\partial t} = -\overrightarrow{\text{rot}}\,\vec{E} + \kappa \vec{\rho} = 0 \quad \Rightarrow \quad \overrightarrow{\text{rot}}\,\vec{E} = \kappa \vec{\rho} \quad \Rightarrow \quad \overrightarrow{\text{rot}}\,\vec{D} = \varepsilon_0 \kappa \vec{\rho} \tag{59}$$

so that the density $\vec{\rho}$ of «vector electric charges» would be the source of a rotational electric field $\vec{E}$ and a rotational electric field of displacement $\vec{D}$, just as the scalar density $\rho$ of electric charges is the source of a divergent electric field of displacement $\vec{D}$

$$\begin{cases} \text{div}\,\vec{D} = \rho \\ \overrightarrow{\text{rot}}\,\vec{D} = \varepsilon_0 \kappa \vec{\rho} \end{cases} \tag{60}$$

If we now compare the coefficients of both theories we obtain the following analogies

$$\varepsilon_0 \Leftrightarrow \frac{1}{2(K_2 + K_3)} \quad \text{and} \quad \kappa \Leftrightarrow 2K_2 \quad \Rightarrow \quad \varepsilon_0 \kappa \Leftrightarrow \frac{K_2}{K_2 + K_3} \tag{61}$$

However the experimental observations of electromagnetism have never shown the existence of such "vector electric charges". Indeed, two reasons can be invoked to explain this state of affairs:

1) the "vector electric charges" simply do NOT exist, which would mean in the case of the analogy with the lattice deformation that the edge dislocations do not exist ($\vec{\lambda} = 0$),

2) the coefficient $\varepsilon_0 \kappa$ is null, or very small, so that we do not observe the presence of these «vector electric charges»

$$|\varepsilon_0 \kappa| \ll 1 \quad \Leftrightarrow \quad \left|\frac{K_2}{K_2 + K_3}\right| \ll 1 \tag{62}$$

which would mean in the case of the analogy with the lattice deformation that the energy of deformation by shear strains is null ($K_2 = 0$), or much more smaller than the energy of deformation by local rotations ($K_2 \ll K_3$).

*- on the generalization of the analogy to other modern physics theories:* the theory of solid lattice deformation developed here in Euler's coordinates is actually a much more complex theory than the classical electromagnetism, since it stems from a tensor theory, which can be reduced to a vector theory by contraction on the tensor indices. Considering the tensor aspect of solid lattice deformation theory, and by relaxing the more restrictive hypothesis (the non-divergent deformations), the analogy could become particularly interesting and fruitful [22,23].

For instance, it is shown in [8] that it is possible to calculate the resting energy $E_0$ of the dislocations, which corresponds to *the elastic energy stored in the lattice* by their presence and to their kinetic energy $E_{cin}$, meaning *the kinetic energy of the lattice particles mobilized by their movement.* This allows to assign to the dislocations *a virtual inertial mass* $M_0$ which satisfies relations similar to the equation $E_0 = M_0 c^2$ of the Einstein special relativity, but which is obtained in this case through purely classical calculations, without using relativity principles.

Moreover, the topological singularities within the lattice (dislocations and disclinations) with their respective charges responsible for the plastic distortions and contortions of the lattice, are also submitted at high velocities (velocities near the transversal wave celerity) to *a relativistic dynamics* within the lattice, and satisfied *the Lorentz transformations,* due to the Maxwell's equations set governing the shear strains and the local rotations of the massive elastic lattice [8,22]. From this point of view, the relativistic dynamics of the topological singularities inside the lattice is a direct consequence of the purely classical newtonian dynamics of the elastic lattice in the absolute frame of the external observer. And this means also that Lorentz transformations and the rules of special relativity *are not an exclusive property of electromagnetic fields and charges*, but a behavior that can appears in different physical systems.

It also appears in the theory developed in [8,22] that the tensor aspect of the distortion fields at short distances of a localized topological singularities cluster formed by one or more dislocation or disclination loops can be neglected at great distances of the cluster, because the distortion fields can then be completely described by only two vector fields, *the vector field of rotation by torsion and the vector field of curvature by flexion*, associated respectively to the only two scalar charges of the cluster, its *scalar rotation charge* $Q_\lambda$ and its *scalar curvature charge* $Q_\theta$

$$\begin{cases} Q_\lambda = \iiint_{cluster} \lambda \, dV = \iiint_{cluster} \mathrm{div}\,\vec{\omega} \, dV = \oiint_{cluster} \vec{\omega} \, d\vec{S} \\ Q_\theta = \iiint_{cluster} \theta \, dV = \iiint_{cluster} \mathrm{div}\,\vec{\lambda} \, dV = \oiint_{cluster} \vec{\lambda}^{div} \, d\vec{S} \end{cases} \quad (63)$$

For example, a linear string of dislocation or disclination can be closed on itself, forming *a loop* inside the lattice [8]. It appears two very interesting cases of such loops, presenting pure scalar rotation charge or pure scalar curvature charge:

- *the twist disclination loop*, bordered by *a pseudo-screw dislocation,* which corresponds to *a pure localized rotation charge* $q_\lambda$,
- *the prismatic dislocation loop,* bordered by *an edge dislocation*, which corresponds to *a pure localized curvature charge* $q_\theta$.

The rotation charge $q_\lambda$ of a twist disclination loop becomes the perfect analogue of *an electric charge* in the Maxwell's equations, as for example *the electron*.

Concerning the curvature charge $q_\theta$ of a prismatic dislocation loop, it has without doubt a role to play in an analogy with the gravitation theory [22,23], even if such a charge does not exist in the modern physics theories as gravitation, quantum physics or particles physics.

Consider for example a solid lattice presenting strong spatial and time variations of its expansion $\tau$. In this case, the equations of table 1 are no more usable as they are deduced in the special case of constant and homogeneous expansion of the lattice, and one has to use the complete Newton's equation *(32)* in order to describe the dynamics of the lattice. The Newton's equation *(32)* can be combined with the expression of the curvature vector $\vec{\chi} = -\overrightarrow{\mathrm{rot}}\,\vec{\omega} + \vec{\lambda}$ of the lattice, and one can write the following equation

$$\underbrace{2(K_2 + K_3)\vec{\chi} + (4K_2/3 + 2K_1)\overrightarrow{\mathrm{grad}}\,\tau}_{generalized\ curvature\ vector\ of\ the\ lattice} = \underbrace{n\frac{d\vec{p}}{dt} - \overrightarrow{\mathrm{grad}}\,U^{def}}_{energy-momentum\ of\ the\ lattice} + \underbrace{2K_3\vec{\lambda}}_{density\ of\ flexion\ charges} \quad (64)$$

Except the last term $2K_3\vec{\lambda}$, this expression presents some analogies with the Einstein's gravitation equation $G = 8\pi T$ of General Relativity, in which $G$ is the Einstein curvature tensor *(Einstein tensor)*, which is expressed in terms of the *Ricci curvature tensor* $G_{\mu\nu} = R_{\mu\nu} - g_{\mu\nu}R/2$, corresponding to a certain part of *the tensor of Riemann* which describes the curvatures of space-time, and in which the tensor $T$ is a «geometrical objet» called *the tensor of energy-momentum* (stress-energy tensor) which characterizes the matter contained in the space. In the case of the equations of field of Einstein, we should also note that the tensor of energy-momentum is a tensor with null divergence $\vec{\nabla} \bullet T = 0$, which translates the conservation of energy and momentum. This equation $\vec{\nabla} \bullet T = 0$ represents in fact the equation of movement in General Relativity.

In the case of the solid lattice deformation in Euler's coordinates, the divergence of relation *(64)* furnishes the following equation

$$\text{div}\left[\left(n\frac{d\vec{p}}{dt} - \overrightarrow{\text{grad}}\,U^{déf}\right) - \left(4K_2/3 + 2K_1\right)\overrightarrow{\text{grad}}\,\tau\right] = 2K_2\theta \qquad (65)$$

which is nothing else than the divergence of the second Newton's equation *(33b)* describing the divergent part of the solid lattice deformation. If one supposes that $\theta = 0$, equation *(65)* becomes the analog of equation $\vec{\nabla} \bullet T = 0$ of general relativity. As an analogy with *the density* $\theta = \text{div}\,\vec{\chi} = \text{div}\,\vec{\lambda}$ *of curvature charges* does not exist in modern physics theories as gravitation, quantum physics or particle physics, one can wonder whether it might be possible to imagine *a new kind of curvature charges* in these theories [22,23].

As reviewed in [20,21], the existence of analogies between the theories of continuum mechanics and solid defects and the theories of electromagnetism, special relativity and gravitation has already been the subject of several publications, from which the most famous are certainly those of Kröner [4,5]. But none of these publications has gone as far as the eulerian approach published in [8,22] concerning the analogies with electromagnetism in the case of non-divergent lattice deformations and with gravitation in the case of divergent lattice deformations, as well as other analogies with all modern physics theories, as explained in [23].

## Conclusion

The analogy presented in this paper between the equations of a non-divergent deformation of an isotropic solid lattice in Euler's coordinates and the Maxwell's equations of electromagnetism is complete, because it is generalized to the two Maxwell's equation couples as well as to *the diverse phenomena of dielectric polarization and magnetization of matter*, just as to *the electric charges and the electric currents*. This implies that the set of Maxwell's equations is *a "model"* in the sense that it can describe different physical systems, and not only the electromagnetism phenomenon.

This work [8] opens the way to develop other analogies between the eulerian solid lattice deformation theory and the modern physics theories [22,23]. It could also and above all be useful to define close links and unifying bridges between the diverse theories of modern physics.

## Acknowledgements

I would like to thank Gianfranco D'Anna, Marc Fleury, Daniele Mari and Willy Benoit for providing valuable input and comments.